\documentclass[structabstract]{aa}

\usepackage{graphicx}
\usepackage{amssymb,amsmath}
\usepackage{natbib}
\usepackage{txfonts}

\def\Msun{ M_\odot}
\def\Rsun{ R_\odot}
\def\Mjup{ M_{\rm J}}

\def\Mp{M_{\rm p}}
\def\Rp{R_{\rm p}}
\def\Ms{M_{\ast}}
\def\Rs{R_{\ast}}
\def\Mearth{ M_\oplus}

\def\Op{\Omega_{\rm p}}
\def\Os{\Omega_{\ast}}
\def\Tp{T_{\rm p}}
\def\Ts{T_\ast}
\def\G{\mathcal{G}}

\def\d{\mathrm{d}}

\def\cm{\mathrm{cm}}

\def\sec{\mathrm{s}}

\def\gr{\mathrm{g}}

\usepackage[normalem]{ulem}
\usepackage{color}
\definecolor{blue}{RGB}{0,0,255}
\definecolor{red}{RGB}{255,0,0}
\definecolor{green}{RGB}{0,200,0}
\definecolor{black}{RGB}{0,0,0}

\begin{document}

\title{Effect of the stellar spin history on the tidal evolution of close-in planets} 
   \subtitle{ }

   \author{ Emeline Bolmont \inst{1,2}
          \and Sean N. Raymond \inst{1,2} 
          \and Jeremy Leconte \inst{3} \and Sean P. Matt \inst{4,5}
                }

   \institute{Univ. Bordeaux, LAB, UMR 5804, F-33270, Floirac, France
   			 \and CNRS, LAB, UMR 5804, F-33270, Floirac, France
         \and Laboratoire de M\'et\'eorologie Dynamique, Institut Pierre Simon Laplace, Paris, France
         \and Laboratoire AIM Paris-Saclay, CEA/Irfu Universit\'e Paris-Diderot CNRS/INSU, 91191 Gif-sur-Yvette, France 	\and	NASA Ames Research Center, M.S. 245-6, Moffett Field, CA 94035-1000, USA}

   \date{Received xxx ; accepted xxx}


  \abstract
{The spin rate of stars evolves substantially during their lifetime, due to the evolution of their internal structure and to external torques arising from the interaction of stars with their environments and stellar winds.}
{We investigate how the evolution of the stellar spin rate affects, and is affected by, planets in close orbits, via star-planet tidal interactions.}
{We used a standard equilibrium tidal model to compute the orbital evolution of single planets orbiting both Sun-like stars and very low-mass stars (0.1 $\Msun$).  We tested two stellar spin evolution profiles, one with fast initial rotation (1.2 day rotation period) and one with slow initial rotation (8 day period).  We tested the effect of varying the stellar and planetary dissipation and the planet's mass and initial orbital radius. }
{For Sun-like stars the different tidal evolution between initially rapidly and slowly rotating stars is only evident for extremely close-in gas giants orbiting highly dissipative stars.  However, for very low mass stars the effect of initial rotation of the star on the planet's evolution is apparent for less massive ($1 \Mearth$) planets and for typical dissipation values.
We also find that planetary evolution can have significant effects on the stellar spin history. In particular, when a planet falls on the star it makes the star spin up.}
{Tidal evolution allows to differentiate the early behaviors of extremely close-in planets orbiting either a rapidly rotating star or a slowly rotating star. The early spin-up of the star allows the close-in planets around fast rotators to survive the early evolution. For planets around M-dwarfs, surviving the early evolution means surviving on Gyr timescales whereas for Sun-like stars the spin-down brings about late mergers of Jupiter planets.  In light of this study, we can say that differentiating between one spin evolution from another given the present position of planets can be very tricky. Unless we can observe some markers of former evolution it is nearly impossible to distinguish the two very different spin profiles, let alone intermediate spin profiles. Though some conclusions can still be drawn from statistical distributions of planets around fully convective M-dwarfs . However, if the tidal evolution brings about a merger late in its history it can also entail a noticeable acceleration of the star in late ages, so that it is possible to have old stars that spin rapidly. This raises the question of better constraining the age of stars.}
		
   
\maketitle
%

\section{Introduction}

The spin rate is an important quantity for the evolution of a star and also for the evolution of any planets orbiting close-in.

The parameter that governs the direction of tidal evolution for a planet orbiting a star (or a satellite orbiting a planet) is the initial semi-major axis with respect to the corotation radius, the orbital radius where the orbital period matches the central body's spin period. For a planet interior to the corotation radius the planet's mean motion is faster than the primary's rotation, so the tidal bulge raised by the planet on the primary lags behind the position of the planet.  The planet feels a drag force that slows it down and causes its orbital radius to shrink, in some cases leading to an eventual merger with the primary. However, for a planet exterior to the corotation radius, the tidal bulge on the star is in advance with respect to the position of the planet and tidal forces push the planet outward.  


The rotational evolution of Sun-like stars can be described in 3 main stages: the pre-main sequence (PMS) stage, the zero-age main sequence (ZAMS) approach and the main sequence (MS) relaxation \citep{Bouvier2008}. During the PMS stage the young stars are observed to have a range of spin periods, typically from a few to $\sim 10$ days, and there is evidence that a highly efficient braking mechanism is at work \citep{Herbst2007}.  It is still not clear what mechanisms are responsible for the observed distribution and the angular momentum loss, but these may be due to the interaction between the star and surrounding accretion disk \citep[e.g.,][]{GhoshLamb1978, Shu1994, MattPudritz2005, Matt2010}, powerful stellar winds \citep{Hartmann1982, Hartmann1989, ToutPringle1992, PaatzCamenzind1996,MattPudritz2005a,MattPudritz2008,Matt2012}, or other processes.  Toward the end of the PMS phase, the fastest stars in the observed distributions appear to spin up in a way consistent with angular momentum conservation, while the rotation rates of the slowest rotators does not appear to change significantly.  Thus, near the ZAMS, the spin period distributions are the widest, typically ranging from a few hours to $\sim 10$ days \citep[e.g.,][]{Bouvier1997, Bouvier2008}.  Once on the main sequence, the stellar structure evolves slowly enough that the torque from ordinary stellar winds becomes important.  Thus, on gigayear timescales, the average spin rates decrease \citep{Skumanich1972}, and the range of observed spin rates narrows.  To bracket the range of observed stellar spin rates, we consider here two populations: initially fast rotators, whose evolution follows the upper envelope of the observed spin rate distributions, and slow rotators that follow the lower envelope.



During the PMS and approach to ZAMS, the observed spin period distributions of M-dwarfs is qualitatively similar to that of Sun-like stars. Observations of young clusters constrain the rotation period of low-mass stars younger than a few $\times 100$~Myr \citep{Stassun1999,Herbst2001,Irwin2008b} but that approach fails for old clusters due to the faintness of old M-dwarfs.  Nonetheless, old slowly-rotating M-dwarfs  have been detected \citep{Benedict1998,Kiraga2007,Charbonneau2009}. 
Contrary to Sun-like stars that are mostly radiative except for a small (in terms of mass) convective region at the surface, very low mass stars ($M_\ast<0.35\Msun$) are entirely convective \citep{ChabrierBaraffe1997}. For Sun-like stars with a radiative core, the interface between the core and convective envelope is thought to be important for the magnetic dynamo, whereas in fully convective low mass stars other mechanisms have to be invoked to explain their observed magnetic activity \citep{ReinersBasri2007}. For example, \citet{ChabrierKuker2006} showed that mean field modeling can produce a $\alpha^2$ dynamo, which creates large-scale nonaxisymmetric fields and \citet{Browning2008} showed that three-dimensional nonlinear magnetohydrodynamic simulations of the interiors of fully convective M-dwarfs can also produce a large-scale dynamo.

%
%

The coupling between stellar spin history and tidal evolution has been studied by \citet{Zahn1994} for close binaries and by \citet{Dobbs-Dixon2004} for short period planets. Individual systems where tidal interactions are thought to have played a role have also been the subject of various studies. \citet{Lin1996} proposed that the planet orbiting 51 Peg stopped its disk-induced inward migration because of its presence outside corotation before the disk dispersal.  Individual systems of the OGLE survey have been studied by \citet{Paetzold2004}. Some studies give constraints for stellar dissipation, such as \citet{CaronePaetzold2007} for OGLE-TR-56b and by \citet{Lanza2011} for the CoRoT-11 system.

In this study, we try to have a more general and systematic approach of the effect of the stellar spin evolution on the tidal evolution of close-in planets. To this end, we couple stellar evolutionary models \citep{Baraffe1998}, wind parametrization \citep{Bouvier2008,Irwin2011} and tidal evolution. We consider two limiting cases for the stellar spin evolution that correspond to a star whose initial rotation is either very fast or very slow.  These different evolutionary paths can be seen in Figure 1 of \citet{Bouvier2008} for Sun-like stars, or in Figures 13 to 15 of \citet{Irwin2011} for M-dwarfs . The slowly rotating stars begin with a rotation period of $8$~days and the fast rotating stars with a period of $1.2$~days. Both fast and slow rotators evolve as explained in \citet{Bouvier1997}, where the loss of angular momentum due to the stellar wind is quantified given different star-dependent parameters among which is the rotation rate of the star. The higher the spin of a star, the more active the star, the stronger the winds and the stronger the braking.

Here we use a standard equilibrium model to study the tidal evolution of planets orbiting stars. Our paper is structured as follows. The tidal and star evolutionary models are briefly discussed in Section \ref{model1}. Some preliminary analysis based on order of magnitude study are made in Section \ref{wtf} before giving the results of the tidal evolution of planets around the two types of stars considered here in Section \ref{Results}. Finally, in Section \ref{Discussion} we discuss the difficulty of linking the results of tidal evolution and observations and the important effect of late mergers on the rotation rate of the stars.


\section{Model description}
\label{model1}

We have developed a model to study the orbital evolution of planets around stars by solving the tidal equations for arbitrary eccentricity and also taking into account the observed spin evolution of stars.

\subsection{Tidal model}
\label{tide_equ}
The tidal model that we used is a re-derivation of the equilibrium tide model of \citet{Hut1981} as in \citet{EKH1998}. We consider both the tide raised by the star on the planet and by the planet on the star. We use the constant time lag model \citep{Leconte2010} and the internal dissipation constant $\sigma$ that was calibrated for giant exoplanets and their host stars by \citet{Hansen2010}. 

\medskip

Taking into account both stellar tide and planetary tide, the secular tidal evolution of the semi-major axis $a$ is given by \citep{Hansen2010}: 
\begin{equation}\label{Hansena}
\begin{split}
\frac{1}{a}\frac{\d a}{\d t} &= -\frac{1}{\Tp}\Big[Na1(e)-\frac{\Op}{n}Na2(e)\Big] \\
& \quad - \frac{1}{\Ts}\Big[Na1(e)-\frac{\Os}{n}Na2(e)\Big],
\end{split}
\end{equation}

where the dissipation timescale $\Ts$ is defined as
\begin{equation}
\label{Tp}
\Ts = \frac{1}{9}\frac{\Ms}{\Mp(\Mp+\Ms)}\frac{a^8}{\Rs^{10}}\frac{1}{\sigma_{\ast}}
\end{equation}
and depends on the stellar mass $\Ms$, its dissipation $\sigma_{\ast}$ and the planet mass $\Mp$. $\Op$ is the planet's rotation frequency, and $n$ is the mean orbital angular frequency. The planet parameters are obtained by switching the $p$ and $\ast$ indices. $Na1(e)$ and $Na2(e)$ are eccentricity-dependent factors, which are valid even for very high eccentricity \citep{Hut1981}:
\begin{align*}
Na1(e) &= \frac{1+31/2e^2+255/8e^4+185/16e^6+85/64e^8}{(1-e^{2})^{15/2}},\\
Na2(e) &= \frac{1+15/2e^2+45/8e^4+5/16e^6}{(1-e^{2})^{6}}.
\end{align*}

The secular tidal equations can be extended to arbitrary obliquity, which has been done in \citet{Leconte2010}.
The equations for the eccentricity and planetary rotation rate can be found in \citet{Bolmont2011}, where tidal evolution was studied for a planet-brown dwarf system.


\subsection{Planets model}

We consider planets with a wide mass range, from $1~\Mearth$ to $5~\Mjup$ ($\Mjup = $~mass of Jupiter). 

For terrestrial planets we use the dissipation values based on Earth's dissipation value. \citet{DeSurgyLaskar1997} inferred the quantity $k_{2,\oplus}\Delta T_{\oplus} = 213$~s from the DE245 data for Earth. $k_{2,p}$ is the Earth's potential Love number of degree $2$, which is a parameter depending on the moment of inertia of the body. It tells us how the body responds to compression ($k_2 = 3/2$ means that the body is an incompressible ideal fluid planet). $\Delta \Tp$ is the constant time-lag. Hansen's $\sigma_{p}$ and the quantity $k_{2,p}\Delta \Tp$ are related through: \begin{flalign} \label{kdtsigma} k_{2,p}\Delta \Tp &= \frac{3}{2}\frac{\Rp^{5}\sigma_{p}}{\G},& \end{flalign}

where $\Rp$ is the planetary radius and $\G$ is the gravitational constant.

The terrestrial planet's compositions are assumed to be an Earth-like mixture of rock and iron, following the mass-radius relation of \citet{Fortney2007}.

For $1~\Mjup$,  $2~\Mjup$ and $5~\Mjup$ planets, we use respectively the values of radius and mass of Jupiter, WASP-33b \citep{Christian2006,CollierCameron2010} and OGLE2-TR-L9 \citep{Snellen2009}. For the dissipation factor, we use \citet{Hansen2010}'s estimates for gas giants : $\sigma_{p} = 2.006 \times 10^{-60}$~g$^{-1}$cm$^{-2}$s$^{-1}$. 

Because the planetary spin synchronization timescale is short compared to the other timescales considered here \citep{Leconte2010,HellerLeconteBarnes2011}, the planet rotation period is fixed to the pseudo-synchronization value at every calculation timestep. In this study, we chose not to treat the evolution of the obliquity of the bodies for simplicity.


\subsection{Stellar evolution}

Sun-like stars and M-dwarfs are self-gravitating objects born from a collapsing dense molecular cloud. In the beginning of their evolution they contract, and when the temperature at their core reaches the value of $\sim 3\times10^6$~K, the PPI nuclear reaction starts and the stars enter the main sequence. The Sun is thought to have entered the MS around a few $10$~Myr after its birth. After $8$~Gyr, the Sun will leave the MS to evolve towards a red giant.

To compute the influence of dissipation into a star, one needs to know the internal structure, mainly the evolution of the radius with time, the moment of inertia (through the gyration radius, which is here considered constant) and the tidal dissipation factor $\sigma_\ast$.
The first two quantities are provided by stars evolutionary models \citep{ChabrierBaraffe1997,Baraffe1998}. However evolution models use unconstrained values for the radius of the star at early ages ($t\lesssim 10^6$yr), so some quantitative uncertainties can arise early in the stellar evolution \citep{Baraffe2002}. However, here we consider evolution after $t_0=5$~Myr for Sun-like stars and after $t_0=8$~Myr for M-dwarfs , so these uncertainties should remain negligible.

We assume that both Sun-like stars rotate as solid bodies, as in \citet{Bouvier1997}, although more recent work has included the effect of internal differential rotation between the radiative core and the convective envelope \citep{Bouvier2008}.  Given that low mass stars are believed to undergo the same kind of stellar wind braking as Sun-like stars \citep{Irwin2011}, we also consider solid rotation \citep{Morin2008}.

In this work, $M_\ast$ is held constant, and the effect of mass loss (through processes like stellar winds) on the internal structure of the star is considered negligible.




\subsubsection{Stellar dissipation}

The stellar dissipation factor is poorly constrained but \citet{Hansen2010} gives estimates for Sun-like stars : $\sigma_{\ast} = 6.4\times10^{-59}~\gr^{-1}\cm^{-2}\sec^{-1} \overline\sigma_{\ast}$ where $\overline\sigma_{\ast} = 7.8\times10^{-8}$ . Thus for Sun-like stars, the dissipation factor is $\sigma_{\ast} = 4.992 \times 10^{-66}~\gr^{-1}\cm^{-2}\sec^{-1}$. 

Like a brown dwarf, a $0.1 \Msun$ star is fully convective so we expect that the dissipation mechanisms within the $0.1 \Msun$ star should be closer to a brown dwarf than a Sun-like star. Thus, we use a dissipation factor of $\sigma_{\ast,dM} = 2.006 \times 10^{-60}~\gr^{-1}\cm^{-2}\sec^{-1}$ for the  $0.1 \Msun$ stars in our calculations \citep{Hansen2010}. No scaling of the M-dwarfs  dissipation factor has been performed compared to the brown dwarfs value. In this work, we varied the stellar dissipation factor by a few orders of magnitude so that the real value for M-dwarfs  is in the considered range. This dissipation is much larger than for a Sun-mass star, and as we will see in subsection \ref{orderm}, this has implications for tidal evolution. 
Stars more massive than $0.35\Msun$ that have radiative zones may be less dissipative and more similar to a Sun-like star.



\subsubsection{Rotational angular velocity}

Our calculations begin during the stellar PMS.  The evolution of the observed spin distributions of PMS stars have often been parameterized in terms of a ``disk locking'' scenario \citep[e.g.,][]{Bouvier1997, Rebull2004, Rebull2006,Edwards1993,ChoiHerbst1996}, in which the spin period of the star is assumed to remain constant at a specified ``initial'' rate, for some specified amount of time (hypothesized to be associated with the dissipation of the disk).  For simplicity, we adopt this basic picture and start our calculations at the point of ``disk dispersal,'' after which the stellar angular momentum evolution is computed according to physical equations.

A primary goal of the present work is to determine how different stellar spin histories influence the star-planet tidal interaction.  To this end, we consider two different spin evolution tracks that approximately follow the fast and slow envelopes of the observed stellar spin distributions, and we do this for two different stellar masses, 0.1 and 1$\Msun$.  In order to describe these tracks in a physically self-consistent way, we adopt a simplified model for the stellar spin, adapted from \citet{Bouvier1997} for solar mass stars and from \citet{Irwin2011} for 0.1$\Msun$ stars.

For both $0.1$ and $1\Msun$ stars we add the effect of tides to the formula of \citet{Bouvier1997} for the loss of angular momentum due to the stellar winds (based on the formulae of \citet{Kawaler1988} and \citet{MacGregorBrenner1991}). The expression for the angular momentum loss rate is: 
\begin{flalign}\label{truc}
  \frac{1}{J}\frac{\d J}{\d t} &= \frac{-1}{J}K\Omega_\ast^\alpha \omega_{sat}^{3-\alpha}\left(\frac{R_{\ast}}{\Rsun}\right)^{1/2}\left(\frac{M_{\ast}}{\Msun}\right)^{-1/2} \\
 & + \frac{1}{J}\frac{h}{2T_\ast}\left[ No1(e)-\frac{\Omega_{\ast}}{n}No2(e)\right], 
\end{flalign}
where $h$ is the orbital angular momentum, $n$ is the mean orbital angular frequency, $T_\ast$ is the stellar dissipation timescale, and the functions $No1$ and $No2$ are defined as:
\begin{align*}
No1(e) &= \frac{1+15/2e^2+45/8e^4+5/16e^6}{(1-e^{2})^{13/2}},\\
No2(e) &= \frac{1+3e^2+3/8e^4}{(1-e^{2})^{5}}.
\end{align*}
 
Here $K$, and $\omega_{sat}$ are parameters of the model from \citet{Bouvier1997}. To reproduce the present rotation of the Sun, we use the value of $K=1.6 \times 10^{47}$~cgs and $\omega_{sat} = 14~\Omega_\odot$. \citet{Bouvier1997} showed that for fast rotators ($\Omega_\ast>\omega_{sat}$), $\alpha = 1$ and for slow rotators ($\Omega_\ast<\omega_{sat}$), $\alpha = 3$.

\citet{Irwin2011} proposed various parameters to reproduce the observational data for stars with masses $0.1<M/\Msun \leq 0.35$, and the spin evolution they simulated was calculated for a star of mass $0.25\Msun$. They also noted that the M-dwarf fast and slow rotators could not be fit with a single value of the parameter $K$. We consider here that the spin evolution of $0.1 \Msun$ stars can be described with the same parameters as $0.25 \Msun$ stars, and we note that this assumption can explain the differences between the curves we show below and the curves of \citet{Irwin2011}.

In this work, we used the values: 
\begin{equation*}
\begin{cases}
 \omega_{sat} = 0.65~\Omega_\odot, & \\
 K_{fast} = 2.03\times 10^{45}~\text{cgs}, & \text{for initially fast rotators} \\
 K_{slow} = 8.0\times 10^{45}~\text{cgs}, &  \text{for initially slow rotators}
\end{cases}
\end{equation*}

These values allows us to reproduce the values of the spin of the star at $t = 5$~Gyr in the two extreme trends seen in Fig. 14 of \citet{Irwin2011}.

	\begin{figure}[h!]
	\begin{center}
	\includegraphics[width=9cm]{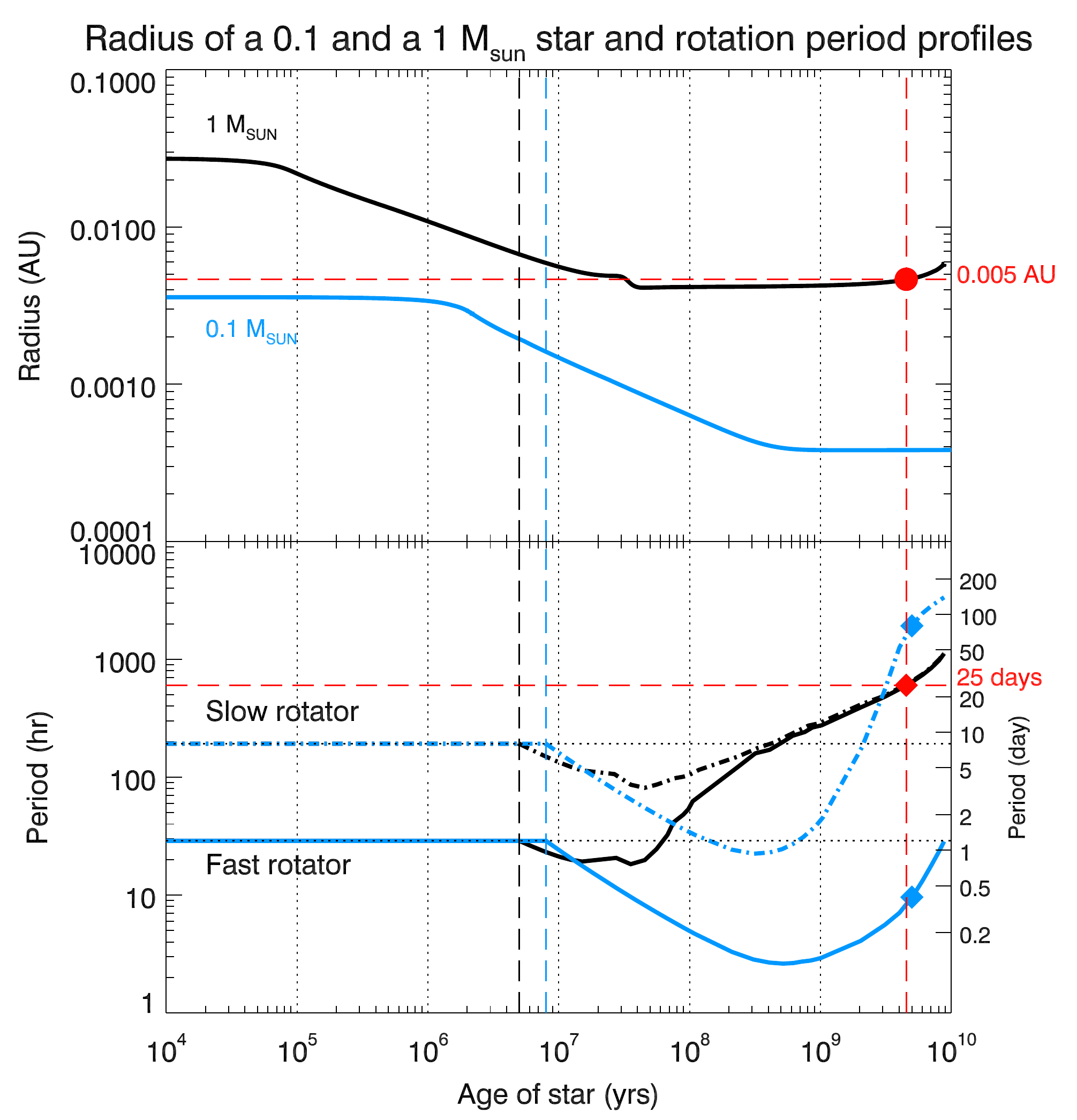}
	\caption{Radius (top panel) and spin (bottom panel) evolution of a $0.1 \Msun$ and a $1 \Msun$ star.  In the bottom panel, the full and dashed dotted blue lines represent respectively the evolution of the rotation period of an initially fast rotating star and an initially slow rotating star with no planet.  As in the top panel, the blue curves correspond to the $0.1 \Msun$ star and the black curves to the Solar-mass star. The blue diamonds in the bottom panel correspond to the values of the spin at $5$~Gyr for the two extreme trends of \citet{Irwin2011}. The vertical dashed lines represent $t_0$ for the two stellar masses.}
	\label{initial_cond}
	\end{center}
	\end{figure}

Fig. \ref{initial_cond} shows $0.1$ and $1 \Msun$ star radius evolution track as well as the different spin period profiles.  For both stellar masses, the slow rotators have an initial period of $P_{\ast 0} = 8$~days and fast rotators  have an initial period of $P_{\ast 0} = 1.2$~days, as in \citet{Bouvier2008}.

Finally, the expression for the stellar rotation is : 
\begin{flalign}\label{truc1}
\Omega_\ast(t) &= \Omega_\ast(t_0) \times\frac{rg2_{\ast}(t_0)}{rg2_{\ast}(t)}\left(\frac{R_{\ast}(t_0)}{R_{\ast}(t)}\right)^2  \\
 & \times \mathrm{exp}\left(\int_{t_0}^{t}f_{tides} \d t\right)\times \mathrm{exp}\left(\int_{t_0}^{t}f_{wind} \d t\right), 
\end{flalign}
where $t_0$ corresponds to the time of disk dispersal, and $f_{tides}$ is given by : 
\begin{equation}\label{Hansenosftides}
\begin{split}
f_{tides} & = \frac{1}{\Omega_{\ast}}\frac{\d\Omega_{\ast}}{\d t}\Big|_{R_{\ast}=cst, rg2_{\ast}=cst} \\
&= \frac{\gamma_{\ast}}{2T_{\ast}}\Big[No1(e)-\frac{\Omega_{\ast}}{n}No2(e)\Big].
\end{split}
\end{equation}
Here, $\gamma_{\ast}=\frac{h}{I_\ast\Omega_\ast}$ is the ratio of orbital angular momentum $h$ to spin angular momentum and $rg2_{\ast}$ is the square of the parameter $rg_{\ast}$ (which is the radius of gyration of \citet{Hut1981}) which is defined as : $I=\Ms(rg_{\ast}R_{\ast})^2$, where $I$ is the moment of inertia of the star. $f_{wind}$ is given by : 
\begin{equation}
f_{wind} =
\begin{cases}
 -\frac{1}{I_{\ast}}K \Omega_{\ast}^{2}\left(\frac{R_{\ast}}{\Rsun}\right)^{1/2}\left(\frac{M_{\ast}}{\Msun}\right)^{-1/2}, & \text{if }\Omega_\ast<\omega_{sat} \\
 -\frac{1}{I_{\ast}}K \omega_{sat}^{2}\left(\frac{R_{\ast}}{\Rsun}\right)^{1/2}\left(\frac{M_{\ast}}{\Msun}\right)^{-1/2}, & \text{if }\Omega_\ast>\omega_{sat}
\end{cases}
\end{equation}

The integration of the equations of Section \ref{tide_equ} was performed using a fourth order Runge-Kutta integrator with an adaptive timestep routine \citep{Press1992}. The precision of the calculations was chosen such that the final semi-major axis of each integrated system was robust to numerical error at a level of at most one part in $10^3$.

Fig. \ref{initial_cond} shows that after the time of disk dispersal $\tau_{disk}$ (long dashed vertical lines in Fig. \ref{initial_cond}), the star spins up due to contraction. After a few $\times 10^8$~yrs, stellar winds start to efficiently spin down the star.


\section{Order of magnitude analysis}
\label{wtf}


\subsection{Parameters space}


We present results for $0.1$ and $1 \Msun$ stars\footnote{Some simulations were performed with a stellar mass of $0.8 \Msun$, but the results were very similar to what we obtained with the $1 \Msun$ star, so we do not discuss those results.}.  We investigate the evolution of planets with masses of $1\Mearth$, $10 \Mearth$, $1M_{Uranus} (= 14.5 \Mearth)$,  $1\Mjup$, $2\Mjup$ and $5\Mjup$.  For the moment we assume zero initial eccentricity so that only the stellar tide governs the evolution. In Subsection \ref{synchronization} we present the outcome of an extreme case with an initial eccentricity of $0.01$.

We consider that planets begin their tidal evolution at the time of disk dispersal $\tau_{disk}$.  The  planet formation timescale is proportional to the orbital frequency \citep{Safronov1969,Raymond2007} and is thus far shorter than the disk lifetime for close-in orbits, but planet-disk interactions probably overwhelm planet-star tidal interactions during this time.  For Sun-like stars, this time is taken to be : $\tau_{disk} = 5\times 10^6$~yrs, which then corresponds to our initial time $t_0$, and for M-dwarfs , $t_0=\tau_{disk} = 8\times 10^6$~yrs.

We tested values for the stellar dissipation $\sigma_\ast$ between 1 and 1000 times the mean values given by \cite{Hansen2010}.

%
%

\subsection{Order of magnitude and timescales}
\label{orderm}

%

The stellar dissipation timescale is given by equation \ref{Tp}.
This timescale depends on the stellar mass $\Ms$, radius $\Rs$, and tidal dissipation factor $\sigma_\ast$, and the semi-major axis $a$ of the planet.  As $\Rs$ shrinks with time, it is clear that for fixed $a$ the stellar dissipation timescale increases with time, so the stellar tide becomes weaker. Assuming $\Mp \ll \Ms$, the ratio of the dissipation timescale of Sun-like stars ($T_{\odot}$) to M-dwarfs ($T_{{ \rm dM}}$) is: 

\begin{equation}
\frac{T_{\odot}}{T_{{ \rm dM}}}  = \left(\frac{R_{{\rm dM}}}{R_{\odot}}\right)^{10} \frac{\sigma_{{\rm dM}}}{\sigma_{\odot}},
\end{equation}
which gives the following values:
\begin{equation*}
\begin{cases}
 T_{\odot} \approx 10~T_{{ \rm dM}}, & \text{at 8 Myr} \\
 T_{\odot} \approx \frac{1}{1000}~T_{{\rm dM}}, &  \text{at 5 Gyr}.
\end{cases}
\end{equation*}

We can see that comparing the magnitude of the tidal effect around a M-dwarf and a Sun-like star is equivalent to compare their radii and their dissipation factors. At all times, Sun-like stars have a bigger radius than M-dwarfs  but have a smaller dissipation factor. 

At $8$~Myr, the dissipation timescale of M-dwarfs  is shorter than of Sun-like stars so the early tidal evolution will be more important around M-dwarfs  than around Sun-like stars. However, at $5$~Gyr, the dissipation timescale of M-dwarfs  is much larger than of Sun-like stars so the late tidal evolution will be more important around Sun-like stars than around M-dwarfs.

	\begin{figure}[h!]
	\begin{center}
	\includegraphics[width=9cm]{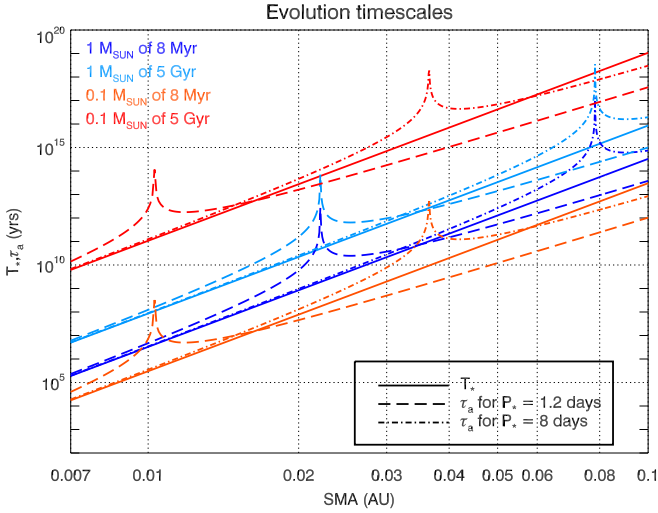}
	\caption{Stellar dissipation timescale $\Ts$ and SMA-evolution timescale $\tau_a$ versus semi-major axis for stars of mass $0.1 \Msun$ and $1 \Msun$ and a planet of Jupiter mass. The stellar dissipation timescale was calculated for mean dissipation factors and for two different star ages: $8$~Myr and $5$~Gyr. The difference between these two times is a difference in the radius of the star, which is smaller at $5$~Gyr than at $8$~Myr.}
	\label{ordermag_bis}
	\end{center}
	\end{figure}

Another important timescale is the semi-major axis evolution timescale (SMA-evolution timescale). Assuming zero orbital eccentricity, it is given by:
\begin{equation}\label{tau_a}
\tau_a  = \left|\frac{a}{\dot{a}}\right|= \left|\Ts\left(1-\frac{\Os}{n}\right)^{-1}\right|.
\end{equation}

In contrast with the stellar dissipation timescale $\Ts$, $\tau_a$ is affected by the spin of the star. $\Ts$ is the limit of $\tau_a$ when $\Omega_\ast \rightarrow 0$.

Figure \ref{ordermag_bis} shows the dependance of the stellar dissipation timescale $\Ts$ and the semi-major axis evolution timescale $\tau_a$ on the semi-major axis $a$ for a system with a Jupiter mass planet, for $M_\ast = 0.1 \Msun$ and $1 \Msun$.  As expected, the farther a planet the weaker the stellar tide.  A Jupiter at $a > 0.05$~AU around a $8$~Myr star of mass $0.1 \Msun$ does not experience any noticeable semi-major axis evolution over $10$~Gyr. Fig. \ref{ordermag_bis} also shows that between $8$~Myr and $5$~Gyr the stellar dissipation timescale increases. At $8$~Myr, $\Ts$ for a $0.1 \Msun$ star is shorter than for a $1 \Msun$ star so the stellar tide will have a stronger effect on the planet orbiting an M-dwarf than a Sun-like star. However, at $5$~Gyr the stellar dissipation timescale of the M-dwarf is longer than $10$~Gyr for planets at $a > 0.007$~AU.  This means that for M-dwarfs the system will ``freeze'' after some time such that the interesting tidal evolution will only occur early in the evolution. For Sun-like stars, the increase of the stellar dissipation timescale is less pronounced so tides still matter for planets closer than $\sim 0.018$~AU at $5$~Gyr.

Concerning the evolution of $\tau_a$ the trends are similar to those for $\Ts$ but a few differences can be seen.  In particular, the curves show a peak feature at a semi-major axis corresponding to the corotation radius.  This is because $\tau_a$ diverges when the semi-major axis is close to the corotation radius: if a planet forms precisely at the corotation radius of a non-evolving star, it will experience no tidal migration. The system is thus perfectly synchronized as the planet - which is already in synchronization - and the star always show each other the same sides. However, the corotation radius is an unstable equilibrium distance because inside the corotation radius - to the left of the peak - the planets migrate inwards and outside the corotation radius - to the right of the peak- the planets migrate outwards. For a system with an evolving star, the stellar tide always causes tidal migration. Fig. \ref{ordermag_bis} also shows that outward migration (taking place further away from the star) always occurs on longer timescales than inward migration (taking place closer in).

Fig. \ref{ordermag_bis} also shows that Jupiter-mass planets closer than $0.02$~AU around Sun-like stars with the mean dissipation value tidally migrate inward. Planets farther than $0.02$~AU experience no noticeable semi-major axis change because $\tau_a$ is so long. In section \ref{Results}, we will show that Sun-like stars need a dissipation factor of $1000 \times \sigma_\ast$ in order to have noticeable tidally-induced changes within $10$~Gyr. In Fig \ref{ordermag_bis}, this is equivalent to lowering the curves corresponding to $1 \Msun$ stars, effectively decreasing the different timescales so that interesting behavior can be observed within the $10$~Gyr timescale we consider here.

We can also consider the timescale of stellar spin evolution $\tau_{\Os}$ (again for zero eccentricity):

\begin{equation}\label{tau_o}
\tau_{\Os}  = \left|\frac{\Os}{\dot{\Omega}_\ast}\right|= \left|\frac{2\Ts}{\gamma_\ast}\left(1-\frac{\Os}{n}\right)^{-1}\right|,
\end{equation}

where $\gamma_\ast = \frac{h}{I_\ast \Os}$ is the ratio of orbital angular momentum $h$ to the spin angular momentum and $I_\ast$ is the moment of inertia of the star. The ratio of the stellar spin synchronization timescale of Sun-like stars to M-dwarfs assuming $\Mp \ll \Ms$ is: 

\begin{equation}
\frac{\tau_{\Omega_\odot}}{\tau_{\Omega_{ \rm dM}}}  = \left(\frac{R_{{\rm dM}}}{R_{\odot}}\right)^{8} \frac{\sigma_{{\rm dM}}}{\sigma_{\odot}}\frac{rg2_\odot}{rg2_{{\rm dM}}}\frac{1-\Omega_{ \rm dM}/n}{1-\Omega_\odot/n},
\end{equation}
where $rg2_i$ is the square of the parameter $rg_i$ (which is the radius of gyration of \citet{Hut1981}).

	\begin{figure}[h!]
	\begin{center}
	\includegraphics[width=9cm]{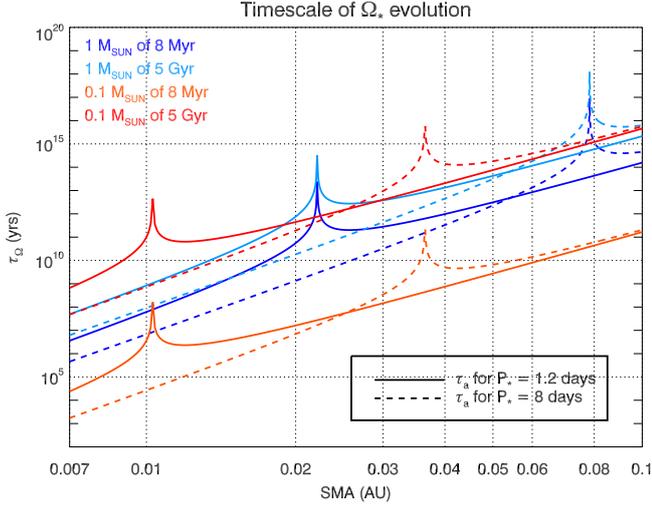}
	\caption{Spin-evolution timescale $\tau_\Omega$ versus SMA for stars of mass $0.1 \Msun$ and $1 \Msun$ and a Jupiter-mass planet. The spin-evolution timescale was calculated for mean dissipation factors and for stellar ages of $8$~Myr and $5$~Gyr.}
	\label{ordermag3}
	\end{center}
	\end{figure}

This timescale depends on the moment of inertia of the star. If the star has a high inertia, the tidal forces need more time to bring the star into synchronization with the orbital frequency. Fig. \ref{ordermag3} shows that the stellar spin evolution timescale is shorter for M-dwarfs than for Sun-like stars at $8$~Myr, and it is longer for M-dwarfs than for Sun-like stars at $5$~Gyr for close-in planets. Due to the strong dependence of tidal effect with respect to the orbital distance, the acceleration of the spin due to a planet inside the corotation radius is faster than its deceleration due to a planet outside the corotation radius. The spin-evolution timescale is short for close-in planets, when the planets fall towards the star there will be an angular momentum transfer between the planets orbit and the spin of the star leading to a noticeable spin-up of the star. 

Stars generally have enough inertia that tidal interactions with a planet require longer than $10$~Gyr to bring the star in synchronization. However, for a very dissipative M-dwarf and a Jupiter mass planet the stellar synchronization timescale is short enough to lead to a perfect synchronized system (as we will discuss in Section \ref{synchronization}). 

The timescale of planet spin evolution is usually much shorter than the stellar spin evolution timescale. That is why in this work we consider the planet is always in pseudo-synchronization -if the eccentricity is non zero- or in synchronization - if the eccentricity is zero.

\section{Results}
\label{Results}

\subsection{Orbital evolution of planets}


For $0.1 \Msun$ stars, the effect of different spin profiles on the orbital evolution of planets is apparent for Earth mass planets and mean dissipations values. For a $1\Mearth$ planet beginning at an initial semi-major axis of $9\times 10^{-3}$~AU, slow rotators systems will tend to make the planet fall on the star whereas fast rotators systems will allow the planet to survive for 10 Gyr. 

\begin{figure}[h!]
	\begin{center}
	\includegraphics[width=9cm]{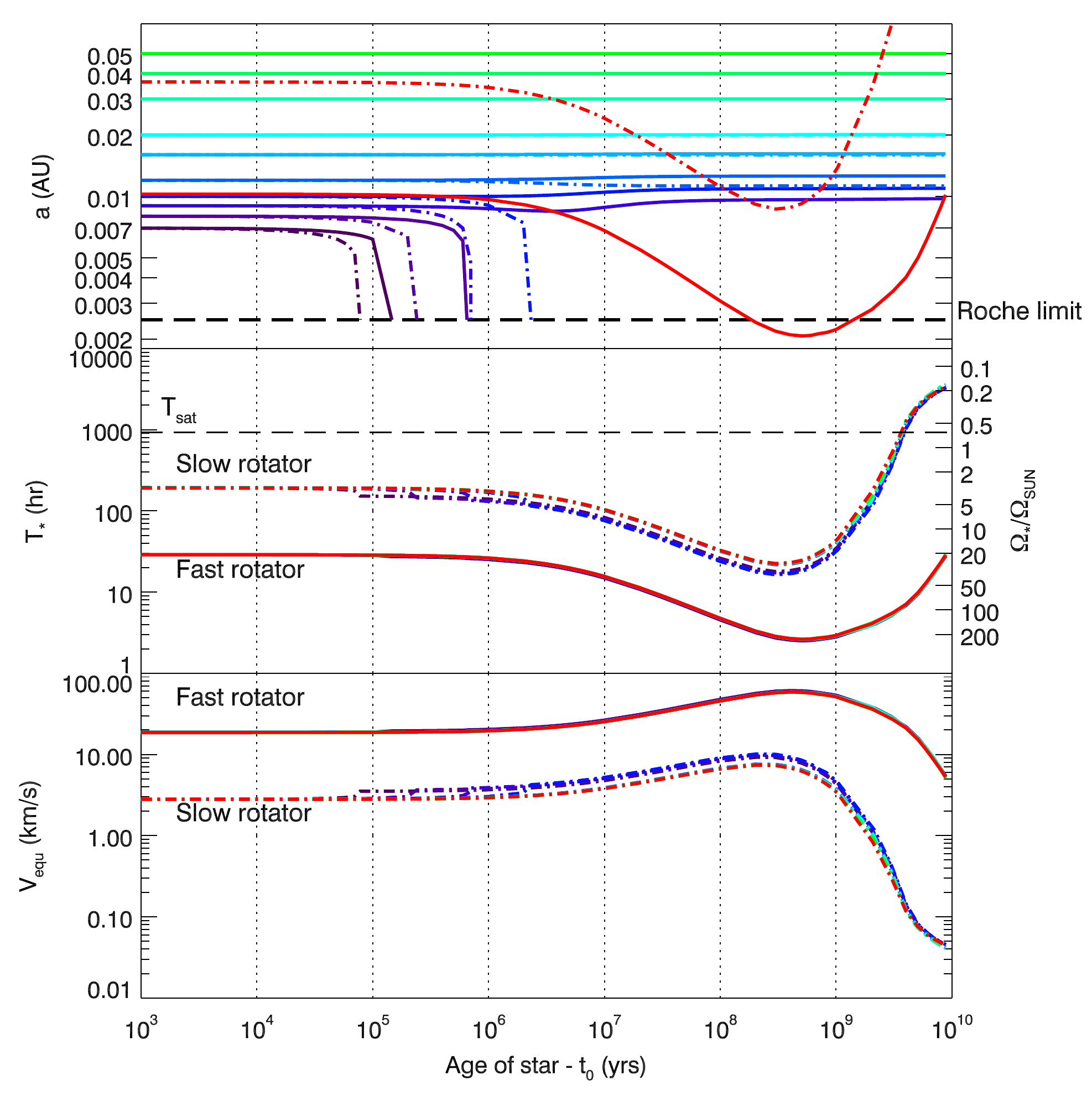}
	\caption{Tidal evolution of $10 \Mearth$ mass planets starting at different initial semi-major axes around either a fast rotating or a slow rotating $0.1 \Msun$ star with mean dissipation factor. Top panel: evolution of the semi-major axis. The full colored lines correspond to fast rotating stars and the dash-dotted lines correspond to slowly rotating stars. The solid and dash-dotted red lines represent the evolution of the corotation radius in both cases if there is no planet. The long black dashes represent the Roche limit. Middle panel: The corresponding stellar rotation evolution (the same line code is used). The black long dashes represent $T_{sat} = 2\pi/\omega_{sat}$. Bottom panel : Equatorial velocity of the star vs time (the same line code is used).}
	\label{10E1}
	\end{center}
	\end{figure}

Interesting effects start to occur for planets of mass higher than $10 \Mearth$. Equation \ref{Tp} shows that the stellar dissipation timescale also depends on the mass of the planet. The more massive the planet, the shorter the dissipation timescale. For $1 \Mearth$ planets, evolution timescales are too big to lead to significant changes in less than a few Gyrs. However for planets of mass higher than $10 \Mearth$, the evolution timescales are compatible with visible changes over a few $10^7$~yrs.  These planets begin to react as they cross the shrinking or expanding corotation radius. Fig. \ref{10E1} shows the results of simulations for $10 \Mearth$ planets orbiting a $0.1 \Msun$ star, using mean dissipation values. A planet beginning at $9\times 10^{-3}$~AU falls on the star in a few hundreds of thousands years if the star is a slow rotator, but survives if the star is a fast rotator. This latter planet experiences a small inward migration before its first crossing with the corotation radius and an outward migration after. For initial semi-major axes larger than $0.02$~AU, the difference between the two spin profiles is negligible. One super Earth, GJ 1214 b, has been detected around a $0.16 \Msun$ star \citep{Charbonneau2009}, so this planet could be experiencing interesting tidal evolution. 

	\begin{figure}[h!]
	\begin{center}
	\includegraphics[width=9cm]{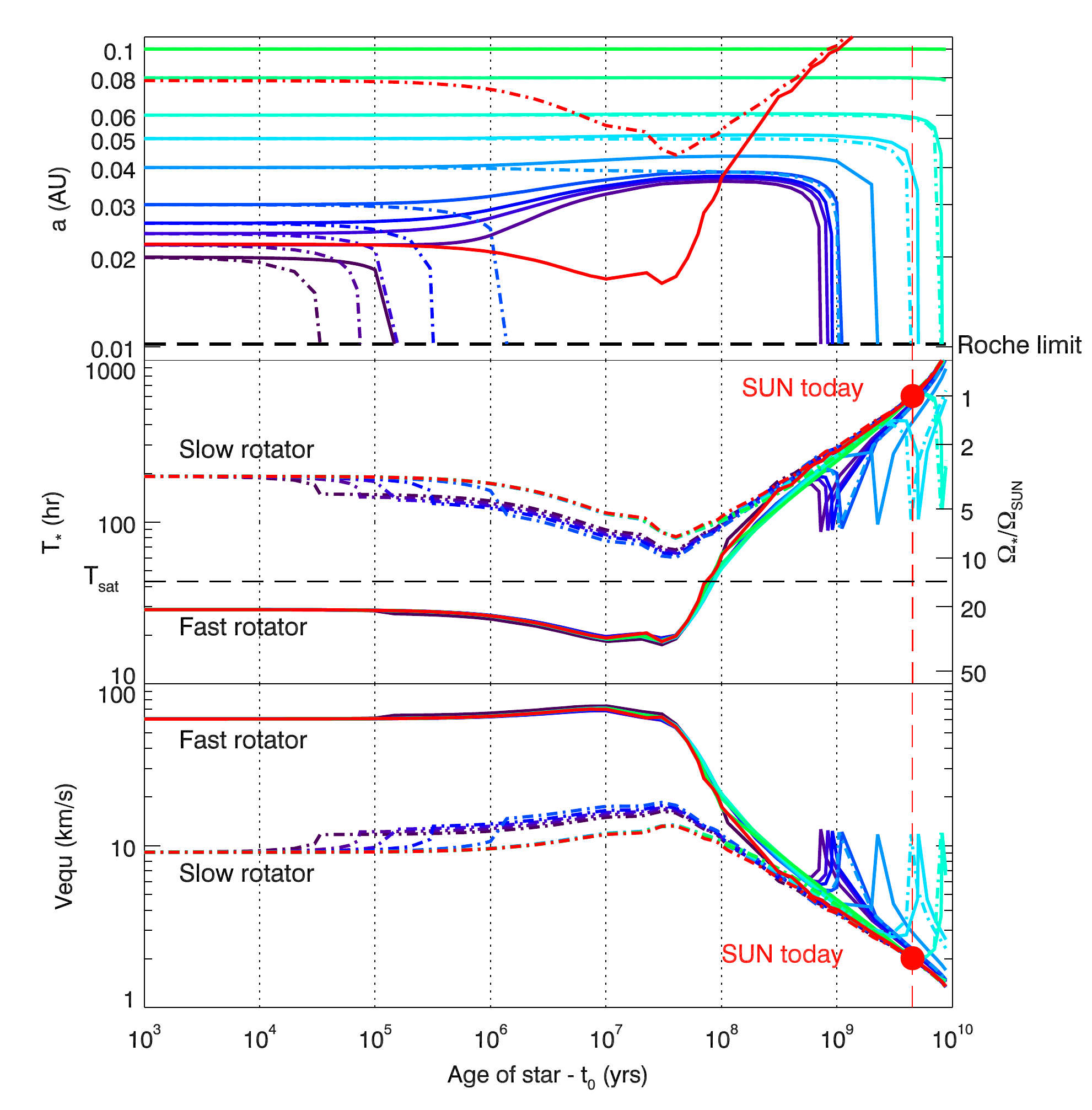}
		\caption{Tidal evolution of Jupiter mass planets of different initial orbital distance around either a fast rotating or a slow rotating $1 \Msun$ star with a large dissipation factor. Top panel: the colored lines represent the semi-major axis of the planets. Solid lines correspond to an initially fast rotating star and dashed dotted lines to an initially slow-rotating star. The red curves represent the corotation radius assuming no planet. Middle panel: evolution of the stellar rotation period. Bottom panel: Evolution of the equatorial velocity of the star. In middle and bottom panel, a red full circle represents the Sun's present rotation, and the red curves correspond to stars with no planets.}

	\label{JUP2}
	\end{center}
	\end{figure}

This effect is more dramatic if the stellar dissipation increased to $1000 \times \sigma_\ast$. The stellar tide contributes in pushing away the planets orbiting a fast rotating star very efficiently and we can find the same behavior as for massive brown dwarfs \citep{Bolmont2011}. For high dissipations and fast rotators, the planets are pushed farther away and for the inner most planets converge towards a given distance. If we considered an equally spaced distribution of planets, the distribution would be more packed at the end of the evolution.



As we saw in Subsection \ref{orderm}, the tidally-induced early semi-major axis evolution occurs on shorter timescale around a $0.1 \Msun$ star than around a $1 \Msun$ star. So to be able to observe any discrepancies between spin profiles for Sun-like stars the stellar dissipation factor must be larger than the mean value. Hereafter, for Sun-like stars  we use a dissipation factor of a thousand times \citet{Hansen2010}'s mean value.

Fig. \ref{JUP2} shows the tidal evolution of Jupiter-mass planets orbiting $1 \Msun$ stars.  Differences can only be seen between fast and slow rotators for very close-in planets. For slow rotators, a planet beginning at $0.03$~AU falls on the star in $\sim 10^6$~yrs, whereas for the fast rotators it falls in $\lesssim10^9$~yrs. 

The evolution of planets around fast rotators is interesting because planets beginning very close to the corotation radius experience an outward migration for a few $\times 10^7$~yrs. Around $t = 50$~Myrs, the star has spun down sufficiently due to stellar winds that these planets cross the outward-drifting corotation radius. After the crossing, these planets migrate back inward and crash onto the star at about$1$~Gyr.  For planets beginning their evolution past $0.05$~AU, the difference between the two spin profiles is negligible.  Planets beginning their evolution at a semi-major axis bigger than $0.08$~AU survive the evolution on $10$~Gyr timescales.

\subsection{Orbital evolution of planets compared with the age of the star}

\begin{figure}[h!]
	\begin{center}
	\includegraphics[width=9cm]{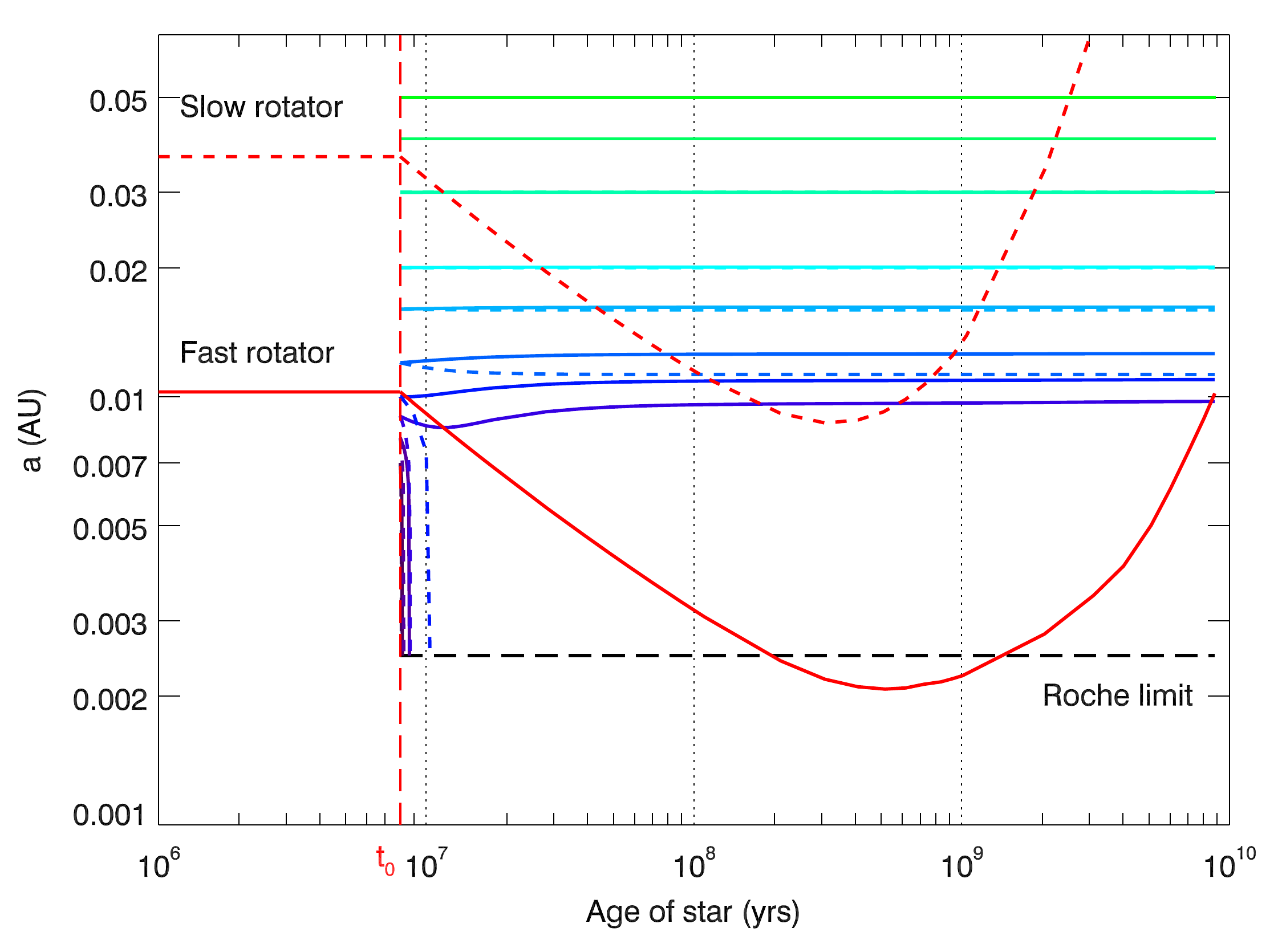}
	\caption{Tidal evolution of a $10 \Mearth$ mass planet starting at different initial semi-major axis around either a fast rotating or a slow rotating $0.1 \Msun$ star. The full colored lines correspond to semi-major axis evolution. The full red line and the dashed red line respectively correspond to the corotation radius of a fast rotating M-dwarf and of a slow rotating M-dwarf with no planet. The black long dashes represent the Roche limit.}
	\label{fulltime}
	\end{center}
	\end{figure}

The time on the x-axis in Figs. \ref{10E1} and \ref{JUP2} starts at $t_0$, thus after $5$-$8$ Myr of stellar evolution, but the stellar evolution timescale may offer a more appropriate measure. Figure \ref{fulltime} shows the orbital evolution of a $10 \Mearth$ planet around a $0.1 \Msun$ star. We can see that the planets which do not survive crash on the star on a timescale much shorter than the age of the star. 

Planets orbiting $0.1 \Msun$ stars undergo similar evolution to planets orbiting brown dwarfs~\citep[see][]{Bolmont2011}. Most of the tidal evolution occurs at early times, when the radius of the stars/brown dwarfs is still large enough that the stellar/brown dwarf tide is strong enough to drive changes in the planets'  semi-major axes. However, the radii of $0.1 \Msun$ stars and brown dwarfs decrease substantially in time such that after a certain interval the stellar/brown dwarf tide becomes weak and the system freezes (see explanation in Subsection \ref{orderm}). 

Thus, we expect that similar conclusions can be drawn for planets around $0.1 \Msun$ stars and brown dwarfs.  A statistical distribution of planets around fully convective M-dwarfs can provide information about their dissipation factors. Indeed, as in \citet{Bolmont2011}, we can make inferences from how densely packed the orbital distribution of close-in planets is; a higher stellar dissipation leading to a more packed distribution of final semi-major axis, and increasing the shortest possible final semi-major axis \citep[see Fig. 19 and 20 of][]{Bolmont2011}.

\medskip

\begin{figure}[h!]
	\begin{center}
	\includegraphics[width=9cm]{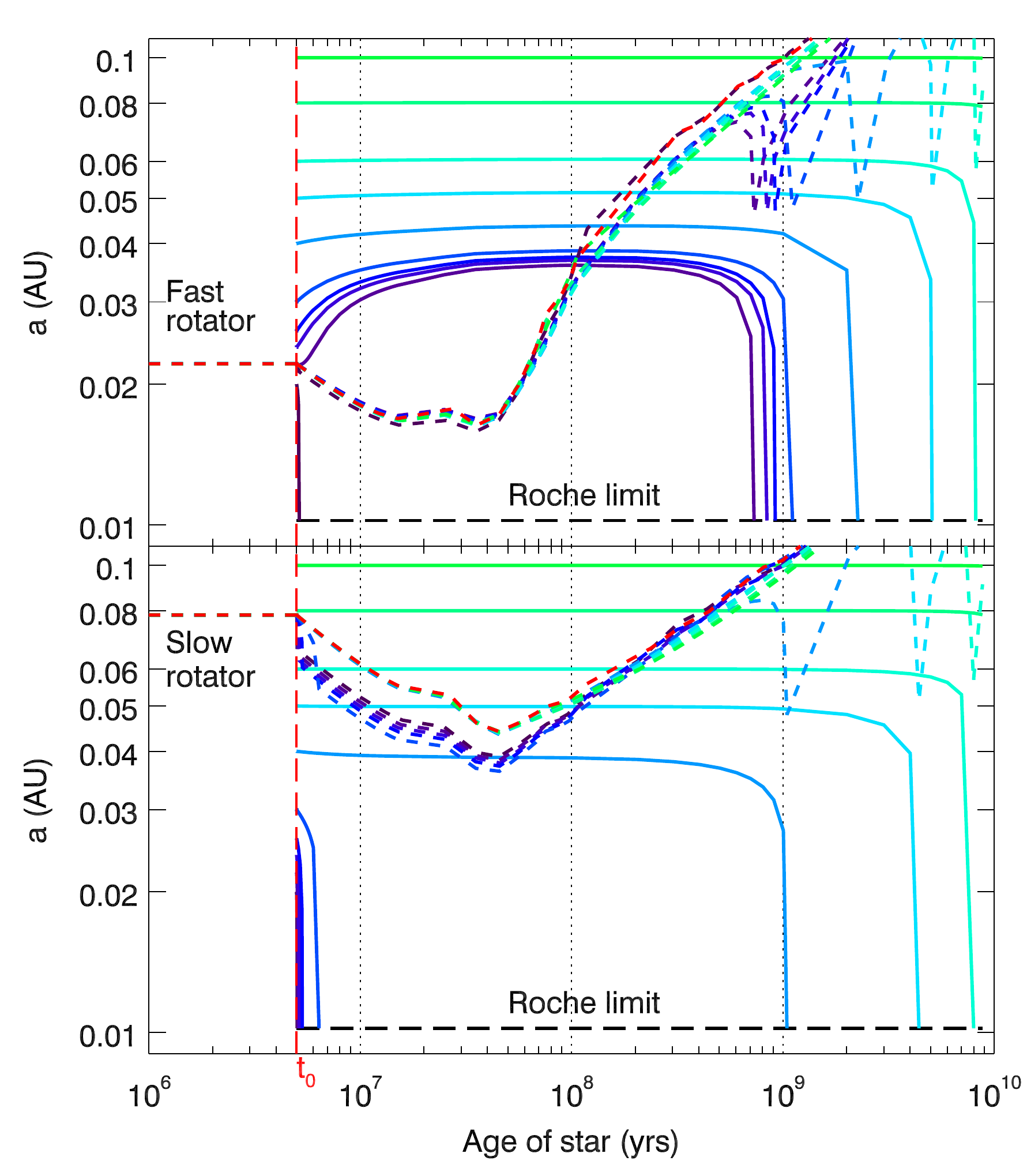}
	\caption{Tidal evolution of a $1 \Mjup$ mass planet starting at different initial semi-major axis around either a fast rotating (top panel) or a slow rotating (bottom panel) $1 \Msun$ star. The full colored lines correspond to semi-major axis evolution. The colored dashed lines correspond to the corotation radius and the red dashed lines correspond to the corotation radius of a star with no planet. The black long dashes represent the Roche limit.}
	\label{fulltime1}
	\end{center}
	\end{figure}

For Jupiter mass planets around $1 \Msun$ stars the situation is different. Planets which survived the early evolution due to the crossing of the shrinking corotation radius still fall on the star on Gyr timescale due to the spin-down of the star. Figure \ref{fulltime1} shows the evolution of Jupiter mass planets around a Sun-like star.

For fast rotators, either the planets fall very quickly compared to the stellar evolution or they fall in more than a few $10^8$~yrs. Contrary to M-dwarfs , the tidal forces at late ages are still important for Sun-like stars (see Subsection \ref{orderm}). So the differences of evolution at early ages due to the difference of initial stellar spin period disappear in the end due to the long term evolution. No conclusions can be drawn as was the case for fully convective M-dwarfs . 

\subsection{Planetary influence on spin evolution}

The spin evolution of a star with no planets is determined by its initial spin, its contraction rate and the efficiency of its stellar wind. However, if a planet is orbiting the star there are angular momentum exchanges between the orbit of the planet and the spin of the star. In particular, if a planet is spiraling inward towards the star, it will have the effect of making the star spin up. 

\subsubsection{Spinning-up and mergers}

\begin{figure}[h!]
	\begin{center}
	\includegraphics[width=9cm]{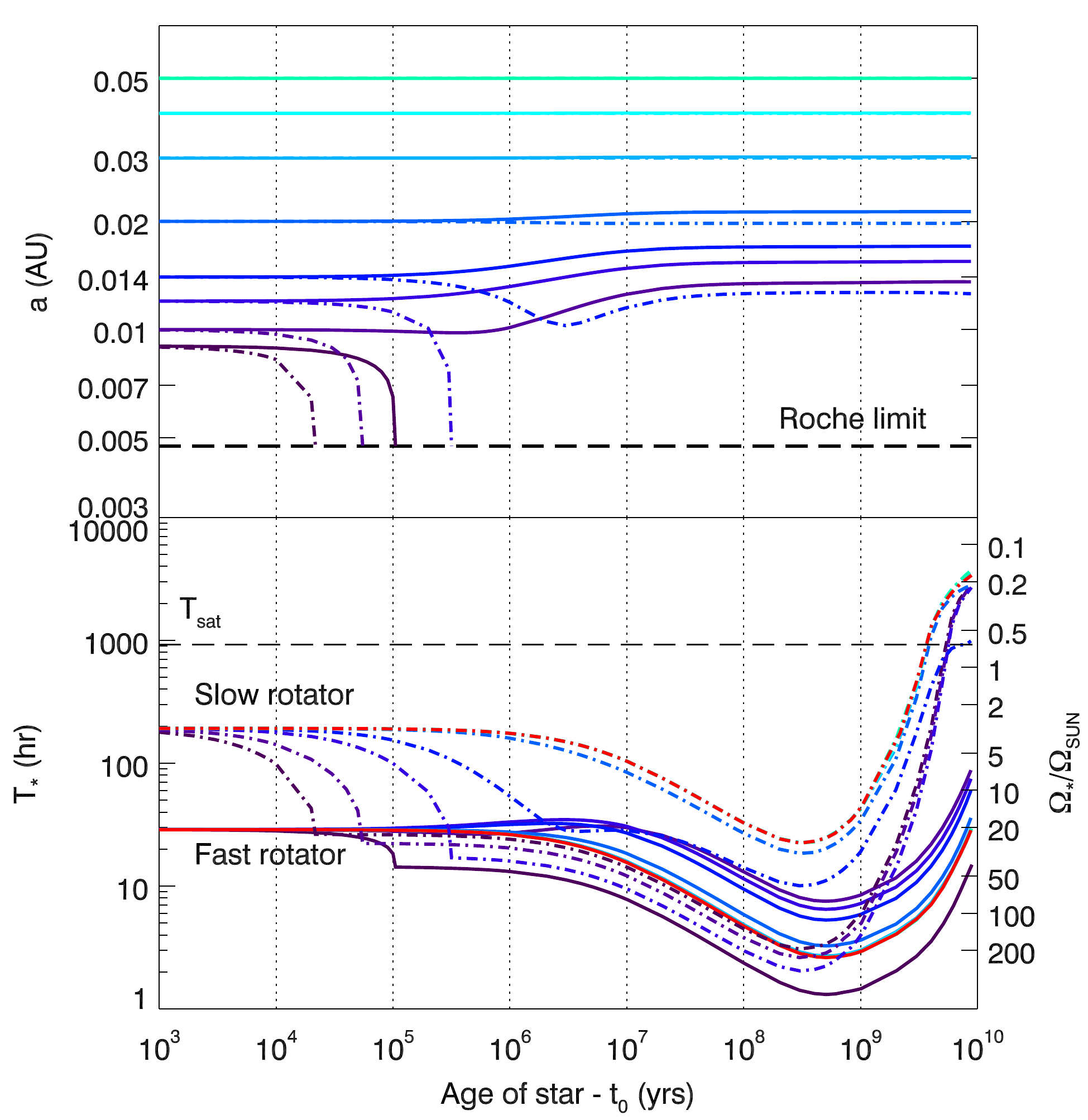}
	\caption{Tidal evolution of a Jupiter mass planet starting at different initial semi-major axis around either fast rotating or a slow rotating $0.1 \Msun$ star. Top panel: evolution of semi-major axis of a a planet around an initially fast rotating M-dwarf (bold solid lines) and around an initially slow rotating M-dwarf (dashed dotted lines). Bottom panel: Rotation period of initially fast rotators (solid lines) and slow rotators (dashed dotted lines). The black long dashes represent $T_{sat} = 2\pi/\omega_{sat}$. The red curves represent the evolution of the spin of the star if there is no planet.}
	\label{JUPm}
	\end{center}
	\end{figure}

Figure \ref{JUPm} shows the evolution of Jupiter-mass planets around either initially fast rotating or an initially slow rotating $0.1 \Msun$ stars, using mean dissipation values.  In \citet{Irwin2011} the time of the dispersal of the gas disk, $\tau_{disk}$, varies with the nature of the star - fast rotator or slow rotator. Here we considered that both fast and slow rotators decoupled from the disk at the same time $t_0 = 8$~Myrs. This way we can compare the effects of tides on the evolution of planets beginning in the same initial conditions - except of course the initial stellar spin. 

As above, we see in Fig. \ref{JUPm} that planets can survive in a wider range of initial semi-major axes around initially fast rotating stars. Compared with Fig. \ref{10E1}, the stellar spin evolution is more strongly modified by planetary migration. When the planet begins spiraling towards the star due to stellar tide, angular momentum is transferred from the planet's orbit to the stellar spin.  Thus, a falling planet can turn a slowly rotating star into a fast rotating star. 

In this work we followed the prescription on \citet{Irwin2011} such that initially fast rotators always have $K_{fast}$ as wind parameter and initially slow rotators always have $K_{slow}$.  However, in the simulations from Fig. \ref{JUPm} some slow rotators actually become fast rotators early in their history.  Given the prescription from  \citet{Irwin2011}, a slow-turned-fast rotator evolves back into a slow rotator due to the difference in wind parameterization.  One can argue that a slow-turned-fast rotator should  rather switch and spin down like a fast rotator instead.  The late spin evolution of such a star would have little effect on any surviving planets because they must lie at larger orbital distances than the planet that perished and also because after a few $\times 10^7$~yrs, the radius of the star is small enough for the tidal effects to be negligible.  

The planet starting at $0.014$~AU around a slowly rotating star (middle panel of Fig. \ref{JUPm}) has an interesting evolution.  The planet starts inside the corotation radius, so the stellar tide pulls it inwards.  Given the planet's large mass, it transfers a significant amount of angular momentum to the star and spins it up.  The corotation radius thus shrinks until it catches up with the planet and inverts the tidal forces on the planet, which then experiences a slow outward migration. At $8\times 10^7$~yrs, the star has spun down due to the stellar winds sufficiently that the planet crosses back inside the corotation radius. For the rest of its evolution the planet migrated slowly inward.

Stellar spin-up when a planet migrates in and falls on the star is also apparent for Sun-like stars. Figure \ref{JUP2} shows delayed peeks in the spin rate  when the star experiences a merger\footnote{\citet{Metzger2012} identified three different outcomes for the merger of a planet into a star, depending on at what orbital distance the Roche lobe of the planet becomes smaller than the actual size of the planet. They also discuss the observational signatures of such events. Some configurations lead to a luminosity peak which could last days or years. However, here we keep a simple description of merger events, and assume that a merger occurs when the planet reaches the Roche limit. Then, all the angular momentum is transferred from the planet's orbit into the spin of the star.
} with a Jupiter-mass planet. When a merger occurs, the star has an excess rotation which disappears on Gyr timescales due to stellar winds. For a system with a planet beginning at $0.06$~AU, the merger occurs after 8~Gyrs, but just after the merger this old star spins five times faster than the Sun today,  corresponding to a equatorial velocity of about $10$~km/s. 

\begin{figure}[h!]
	\begin{center}
	\includegraphics[width=9cm]{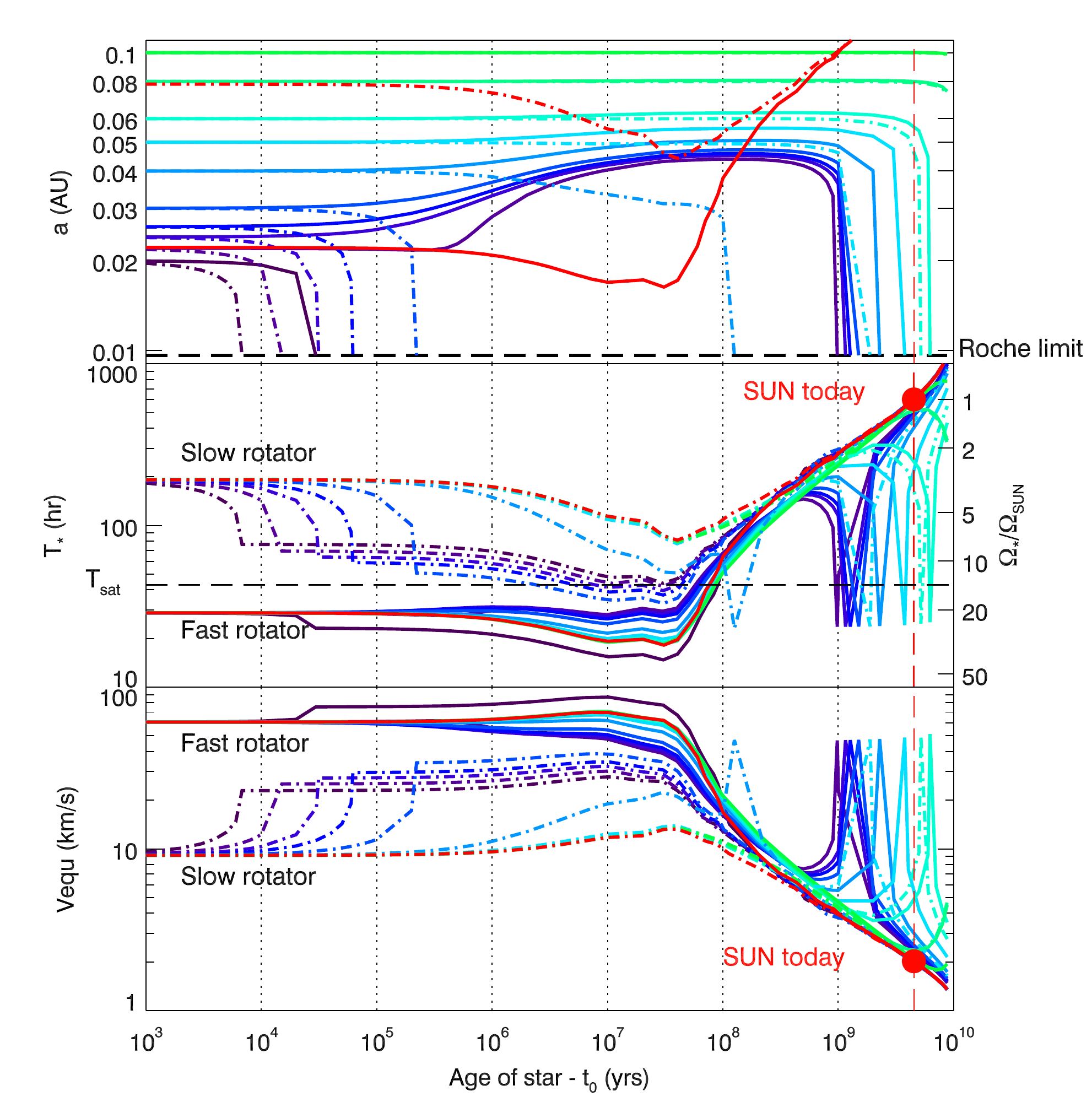}
	\caption{As Figure \ref{JUP2}, but with $5\Mjup$ planets. We can see that the planet being more massive than in Fig. \ref{JUP2} the acceleration of the star is more visible. Some slow rotators even become fast rotators for a few $10$~Myr because of the merger of the planet. When a merger occurs the star is spun up to a rotation rate which is of the order of $20~\Omega_{\odot}$.}
	\label{JUP3}
	\end{center}
	\end{figure}

This effect is stronger for more massive planets. A $5~\Mjup$ planet beginning its evolution at $0.06$~AU falls on the star in about $8$~Gyr. This makes the parent star spin up to more than $20$ times faster than present Sun (see Fig. \ref{JUP3}), with an equatorial velocity of about $60$~km/s. This corresponds approximatively to the fast rotators' initial rotation period.

Unlike close-in planets around $0.1 \Msun$ M-dwarfs , which survived for Gyr timescales if they did not fall in first few Myrs, close-in planets orbiting Sun-like stars that survived the early evolution are still doomed to fall on the star. This difference in behavior can be explained by the fact that M-dwarfs have much smaller radii than Sun-like stars at late ages and thus much weaker stellar tides. For a planet at $0.03$~AU, the semi-major axis evolution timescale $\tau_a$ for a $5$~Gyr old M-dwarf is longer than $10$~Gyr (Fig. \ref{ordermag_bis}), but just $1$~Gyr for a $5$~Gyr old Sun-like star with a dissipation of $1000 \times \sigma_\ast$. M dwarf stellar tides are simply not strong enough to make the planets fall at late times. 


\subsubsection{Stellar synchronization}
\label{synchronization}

We now study a system that displays relatively spectacular tidal evolution.  The system is made up of a Jupiter-mass planet orbiting an initially fast rotating, highly dissipative ($1000 \times \sigma_\ast$) $0.1 \Msun$ star. The planetary dissipation is set to zero to isolate the influence the stellar tide.


	\begin{figure}[h!]
	\begin{center}
	\includegraphics[width=9cm]{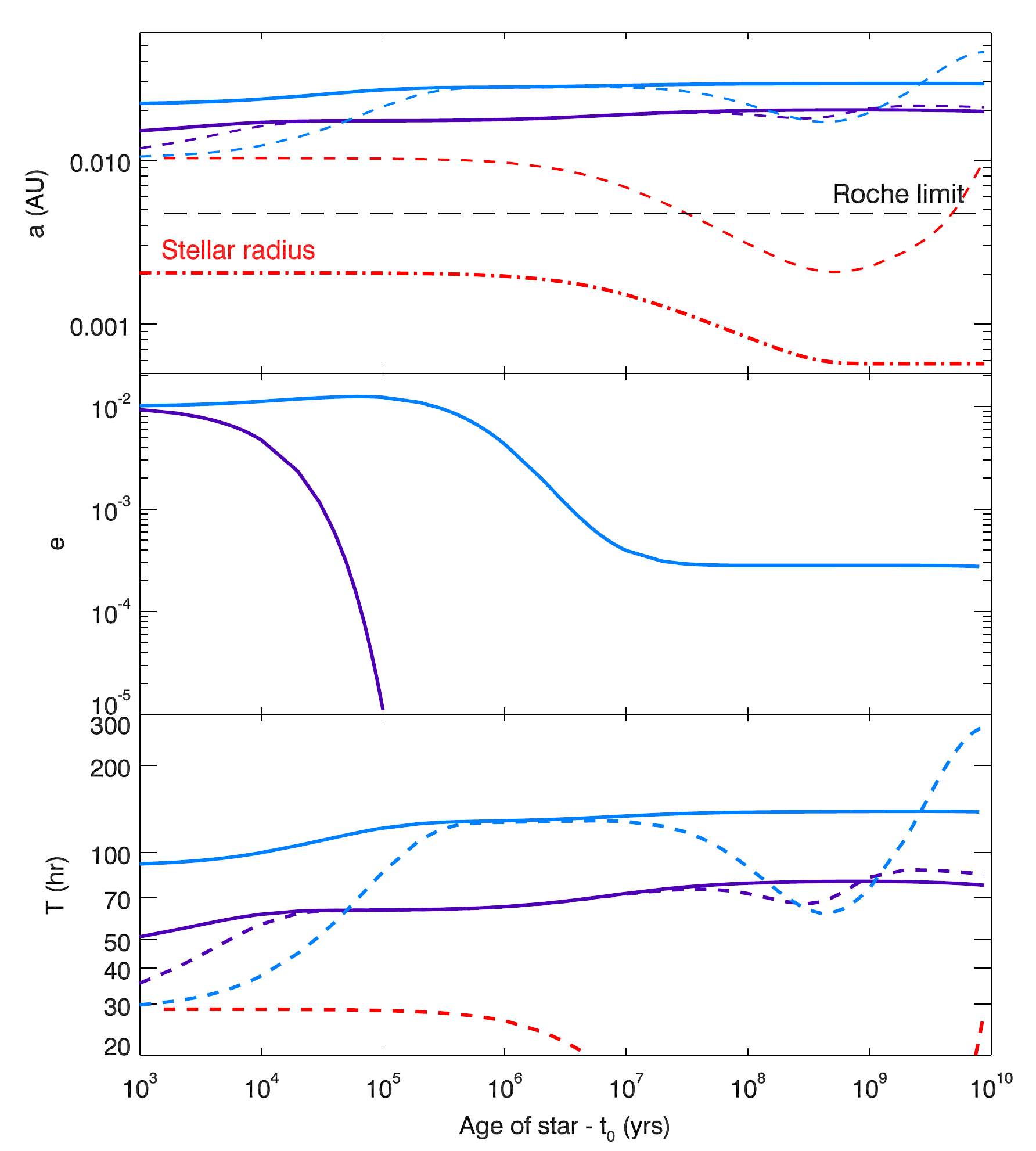}
	\caption{Tidal evolution of a Jupiter mass planet around a fast rotating $0.1 \Msun$ star with a high dissipation factor. Top panel : Evolution of semi-major axis of Jupiter mass planets (full line) beginning at different initial semi-major axis and of the corresponding  corotation distance (dashed line). The red dashed line corresponds to the corotation radius of a star with no planet. Middle panel : Evolution of the eccentricity of the planets. Bottom panel : Evolution of the rotation period of the planets (full line) and of the corresponding star (dashed line). In both cases, the stellar spin initially increases to reach the spin period of the planet (we consider planets in pseudo-synchronization). Later in the evolution either the spin-up due to contraction or the spin-down due to stellar wind leads the spin of the star to differentiate from the spin of the planet.}
	\label{JUP}
	\end{center}
	\end{figure}


Tidal theory states that an equilibrium state is obtained for the two body problem when the orbit is circular, and when both orbit and spins are aligned and synchronized \citep{Hut1980}. In this equilibrium state, both planet and star thus have a spin which is equal to the mean orbital frequency, $\Os = \Omega_p = n$. The planet having a low inertia compared to the star, it will reach synchronization early in the evolution, typically in a few thousand years. However, usually the star because of its higher inertia never reaches synchronization. In most cases, it would require much more than $10$~Gyr for the star to reach synchronization. However, in the example shown in Fig. \ref{JUP}, where the stellar dissipation has been significantly increased, the stellar tide is efficient enough to bring about synchronization of the stellar rotation with the spin of the planet in a few $10^4$~years to a few $10^5$~years.

Fig. \ref{JUP} shows a case. Let us focus on the case of the purple curve which corresponds to a planet beginning at $0.014$~AU. Initially the stellar rotation period is of $1.2$~days, which corresponds to an corotation distance of about $0.01$~AU. In $10^4$~years, the corotation radius moves outward and reached the planet's orbital distance (as can be seen in the bottom panel, the stellar spin period increases to match the planet's orbital period; recall that we assume pseudo-synchronization for planets). After $10^5$~years the stellar tide has made the eccentricity decrease to zero. Indeed, because $\Omega_\ast/n < \frac{18}{11}\frac{Ne1(e)}{Ne2(e)}$ and because of the pseudo synchronization of the planet, this system would be in the region where $\dot{e}<0$ of the phase diagram $\omega/n$ vs $e$ presented in \citet{Leconte2010}. Between $10^5$ years and $10^7$~years, the star and the planet remain in perfect synchronization with a circular orbit and spin synchronization. This is a equilibrium state for the system, or it would be if the star was not evolving in time. Indeed, the stellar radius shrinks which acts to spin-up the star while stellar winds act to spin-down the star, and these effects drive the system away from this equilibrium state. After a few $\times 10^7$~years, contraction spins the star up, and then stellar winds make the star spin down until $1$~Gyr. The planet has crossed the corotation radius and the stellar tide causes it to fall slowly towards the star. The orbital angular momentum of the planet is transferred to the star and spins it slowly back  up.

The other (blue) planet in Fig. \ref{JUP} shows a similar evolution but with two differences.  First, the effect of the stellar tide is weaker because the planet is initially farther out ($0.022$~AU), so synchronization occurs later and for a shorter time. Second, the other physical phenomena influencing the spin evolution of the star kick in earlier. The acceleration phase is much more pronounced for this case than for the previous one. The second difference is that initially,  $\Omega_\ast/n > \frac{18}{11}\frac{Ne1(e)}{Ne2(e)}$, the star is in the phase space region where $\dot{e}>0$, so the stellar tide contributes in increasing the eccentricity. However, the spin of the star decreases faster than the mean orbital angular frequency $n$ so around a few $10^4$~years, the system is back in the region where $\dot{e}<0$ and the stellar tide contributes in decreasing the eccentricity. Between $\sim 6$~Myr and $\sim 2$~Gyr, the planet is outside the corotation radius and $\Omega_\ast/n > \frac{18}{11}\frac{Ne1(e)}{Ne2(e)}$ so the stellar tide contributes in increasing the eccentricity once again. However, in the meantime the radius has shrunk to values such that the stellar tide is negligible and a small eccentricity is kept in the system. When the star starts to spin down due to the stellar wind, the both planet and star are once more in the phase space region where $\dot{e}<0$.

This case represents an interesting case study although it is not likely that the stellar dissipation is strong enough to allow such evolution.

\subsubsection{Implication on stellar age determination}

As we have seen, tidal interactions between a star and a hot Jupiter can bring about significant acceleration of the spin of the star in some cases. This is the same main conclusion as \citet{Pont2009}, who explains that the observed excess rotation of stars with a transiting system can be due to the tidal interaction between the planet and the star.  We note that stars without planets may still bear the traces of their violent pasts due to planet-star mergers.  We can see in Fig. \ref{JUP2} and \ref{JUP3} that while approaching the star, the spin of the star changes continuously to reach a maximum at the merger. 

To determine the age of a star, one technique is ``stellar gyrochronology'' \citep{Barnes2003,Barnes2010,Meibom2011,EpsteinPinsonneault2012}. It consists of measuring the spin of a star, and inferring the age of the star by assuming a spin history. Generally speaking, fast rotating stars are classified as young stars and slow rotating stars are classified as old stars that have been spun down due to stellar winds.

In light of this study, we can see that this method might, in some cases, imply the wrong ages of rapidly rotating stars. A rapidly rotating star could be a young star but it could also be an old star which experienced a merger with a planet. For example, the merger of $5~\Mjup$ planet on an initially fast or slow rotating makes the star spin up almost to the initial rotation period of the group of fast rotators. Thus, old stars that are spun up would be mistaken for young objects.



An old star that underwent a merger with a Jupiter size planet would reproduce several characteristics of young stars: they will be fast rotators, they will have infrared excess due to a hot dust disk and accretion signatures. A merger will also produce extreme UV signatures, soft X-rays and a peak of luminosity depending on the nature of the merger \citep{Metzger2012}.  These stars would most likely be mistaken for young stars.

Gyrochronological ages of fast rotators should be checked against other techniques, in order to verify their youth. Another technique consists of estimating the age of the star by comparing its location on the Herzprung-Russel diagram with theoretical stellar evolution tracks.

\section{Conclusions}
\label{Discussion}

We have shown that the stellar spin history affects the tidal evolution of close-in planets, albeit in a confined region of parameter space.  In cases where the stellar spin history does matter, the difference between two close-in planets -- one orbiting an initially fast-rotating star and the other orbiting an initially slow-rotating star -- comes from an early phase of outward tidal migration for the planet around the fast rotator (caused by the star's closer-in corotation distance).  This phase of outward migration does not occur for the planet orbiting the slow rotator. This outward migration and the decrease of the radius of the star weaken the later tidal evolution and effectively delay or sometimes even prevent the later in-spiralling of planets onto their stars.  At later times both slow and fast rotators spin down due to stellar winds \citep{Skumanich1972,Bouvier1997,Bouvier2008,Irwin2011}, so the orbital history of close-in planets orbiting old stars depends on something that is not directly observable: the stellar spin evolution.

For Sun-like stars the stellar spin history affects tidal evolution only in relatively extreme circumstances.  In particular, the spin history has a strong effect if stars are very dissipative, with dissipation rates $\sigma_\ast$ of about $1000$ times the fiducial value \citep{Hansen2010}.  For strongly dissipative stars, there are differences in tidal evolution for planets orbiting initially slowly- vs. rapidly rotating stars for very close-in ($a \lesssim 0.05$~AU), massive planets ($M_p \gtrsim M_J$).

In contrast, the stellar spin history plays a role in a much wider region of parameter space for $0.1 \Msun$ stars, mainly because these stars are fully convective and so we think they are much more dissipative, with dissipation factors assumed to match those of brown dwarfs.  For these stars the differences in tidal evolution for planets orbiting initially slowly- vs. fast-rotating stars are apparent for mean dissipation values, for planets out to $\sim 0.02$~AU, and for planets with masses as small as $1 \Mearth$.


Low mass stars and Sun-like stars have different dissipation factors and different radii, so the evolution timescales are different and evolve differently (see Figs. \ref{ordermag_bis} and \ref{ordermag3}). The early tidal interaction is stronger for planets around very low mass stars, and the difference between fast rotator profile and slow rotator profile is apparent for planets of one Earth mass with mean dissipation values.  After some time, the system freezes in a given state because the radius of the star has shrunk too much for the tidal evolution to occur on less than $10$~Gyr timescale. For Sun-like stars, the tidal evolution for mean dissipation factor occurs on very long timescales, which is why much more massive planets and higher stellar dissipation rates are needed to produce stellar spin-driven differences in tidal evolution. 


The inward migration of a Jupiter orbiting inside the corotation radius of an initially slow rotating $0.1 \Msun$ star can lead to a significant spinning-up of the star. By the transfer of angular momentum from the planet's orbit, an initially slow rotating star can become a fast rotating star. This effect is more dramatic if the planet actually falls on the star. \citet{Irwin2011} used different wind parametrization for fast or slow rotators so they can infer from the present spin rate if the star was initially fast rotating or not. However we show here that it might not be that straight forward. If the star experienced a merger with a planet it can modify the rotation rate of the star and change the slow rotator into a fast rotator. 

Massive planets orbiting very low-mass stars with high dissipation rates ($\sigma_\ast \times 1000$) can create systems in perfect synchronization where the spin of the star is equal to the spin of the planet (Fig. \ref{JUP}). However, the equilibrium is not stable and the system departs from it as the star spins up due to contraction or spins down due to the stellar winds. This strongly alters the stellar spin profile because the star can be efficiently spun down by a planet initially located outside the corotation radius or spun up by a planet interior to corotation.  

Unfortunately, hot Jupiters around M-dwarfs are extremely rare due to the inefficiency of the planets formation processes around low mass stars \citep{Laughlin2004,IdaLin2005,Kennedy2008}. Only three hot Jupiters are known to exist around stars with masses less than $0.7 \Msun$ \citep{Pepe2004,Hellier2011,Borucki2011,Johnson2012} and none around stars with masses as small as $0.1 \Msun$. Hot Jupiters around very low mass stars remain to be detected.


A statistical distribution of planets around fully convective M-dwarfs  could constrain the tidal dissipation factor $\sigma_\ast$.  Specifically, from the location of the inner edge of the planetary semi-major distribution one can infer a inferior limit for the dissipation factor. The more distant the inner edge, the more dissipative the star. However, in order to draw these conclusions, one needs a good estimate of the stellar ages. 

For Sun-like stars such conclusions cannot be made because, if the dissipation rate is high enough to affect the orbital evolution at early times, significant tidal evolution still takes place at late times, as well. Slow rotators and fast rotators have similar evolutions after a few $10^8$~yrs so the observation of a hot Jupiter orbiting a star of known age and known dissipation would not allow us to infer if the stellar spin history. Indeed, in both cases, different initial semi-major axis can lead to the same observed semi-major axis.  One can imagine trying to infer a planet's orbital evolution from the composition of its atmosphere to know if the planet came from a "cold" region ($0.04$~AU) or a "hot" region ($0.03$~AU). Unfortunately, this exercise is fraught with uncertainties in both the expected atmospheric composition of planets that form at different orbital distances and the tidal parameters ($\Omega_{\ast,0}$, $\sigma_\ast$, the age of observed starsÉ).  

Nonetheless, we emphasize the planets crashing on the star at late ages can entail a significant spin-up of the star and create a population of old fast rotating stars. The spin-up of the star due to a merger has been pointed out in \citet{Levrard2009}, where they found that planets falling on their host star due to tides never reach a tidal equilibrium. \citet{Jackson2009} also addressed the problem of tidally induced mergers and the effect of these mergers on the parent star. They also found that a considerable spin-up is to be expected and also a change in stellar composition. Planet-star mergers thus may confuse stellar age determinations. In general, fast rotators are thought to be young, although we have shown that a merger can lead to old, fast rotating stars that would mimic many of the characteristics of young stars. An independent determination of the age of observed stars is therefore very important, especially for fast rotators.

\begin{acknowledgements} We are grateful to Andrew West for suggesting that we study low-mass stars rather than simply concentrating on the case of Sun-like stars.  We thank Franck Selsis and Franck Hersant for their ideas and support.  We also thank the CNRS's PNP program for funding, and the Conseil Regional d'Aquitaine for help purchasing the computers on which these calculations were performed. SPM acknowledges support by the ERC through grant 207430 STARS2 (http://www.stars2.eu)..  \end{acknowledgements}


%

\end{document}